# Physics-guided machine learning for sim-to-real calibration of NV diamond magnetometers

Jonathan Daniel[†1], Sangwoo Lee[†2], Jesse Hernandez[1], Emanuel Suarez[1], Jin Hee Lee[2], Je-Hyung Kim*[2] and Martin Y. Kim*[1]

*Corresponding author email: jehyungkim@unist.ac.kr; youngmin.kim@csusb.edu

[1]Department of Physics and Astronomy, California State University, San Bernardino, CA 92407 USA
[2]Department of Physics, Ulsan National Institute for Science and Technology, Ulsan 44979, Republic of Korea

## Abstract

Ensemble nitrogen-vacancy centers in diamond enable robust magnetometry in unshielded environments, yet deployment remains bottlenecked by complex calibration and a reliance on external references. Conventional statistical machine learning requires massive training datasets and suffers from severe simulation-to-reality mismatches. To address this, we introduce a physics-guided hybrid machine learning framework that embeds Zeeman splitting directly into the learning pipeline. In an ablation study, our model demonstrates a nearly 3-fold improvement in precision and a 372-fold reduction in mean prediction bias over a purely statistical baseline. Furthermore, anchoring scalable synthetic data generation with just 49 sparse experimental seeds circumvents physical data scarcity while incorporating real-world non-idealities. Evaluated across a hybrid dataset of over 19,000 spectra spanning a 400 G dynamic range, the framework bridges the simulation-to-reality gap and achieves robust scalar field prediction (overall MAE = 4.76 G, $\sigma = \pm$ 9.30 G, and $R^2$ = 0.994), transferring effectively to raw, uncalibrated experimental data. Operating with minimal computational overhead on standard commercial hardware, the model achieves sub-4 ms inference times per spectrum, demonstrating strong viability for real-time edge deployment.

Ensemble nitrogen-vacancy (NV) centers in diamond have matured into a leading solid-state quantum sensing platform due to their atomic-scale spatial resolution[1–3], wide dynamic range[4], and sub-nT sensitivity[4,5], some reaching down to $fT/\sqrt{Hz}$[6] under ambient conditions[7–9]. These properties make them highly appealing for field-ready quantum magnetometers[10–13] deployed on mobile platforms, defense assets, and remote geophysical survey tools[14]. However, implementing these systems out of controlled laboratory environments introduces severe operational challenges. Signal fluctuations in ambient temperature, local strains, microwave power drifts, and unpredictable background noise degrade the quality of Optically Detected Magnetic Resonance (ODMR) data acquisition[15,16]. To maintain metrological accuracy in field applications, modern instruments typically rely on periodic manual recalibration or continuous connectivity to external references like GPS or local networks[14]. For NV magnetometers to be field-deployed in unshielded or GPS-denied field environments, sensors must instead possess an internal mechanism to calibrate assisted by sparse measurements directly from raw, unconditioned experimental data[17].

Conventional calibration methods use non-linear least-squares fitting algorithms to extract spin-resonance frequencies[18,19]. While accurate, they scale poorly when processing low-contrast, asymmetric[16], or noisy signals[20], and they fail completely without precise initial parameters[21]. Alternative strategies, such as real-time numerical fitting solvers and Bayesian methods bypass these challenges and provide high accuracy but introduce severe resource bottlenecks for deployed edge sensors, requiring intensive on-the-fly computational overhead[22].

Artificial neural networks (ANNs) offer a compelling alternative due to their rapid, millisecond execution times and global convergence capabilities[17–19]. Further improvements have been recently reported using machine learning on NV magnetometer measurement system with a sequential Bayesian experiment design[23] and Bayesian inference with a graybox model[24]. Once trained, a neural network condenses vast physical parameter trajectories into a static, highly optimized weight matrix. This exceptional data compression[25] makes neural architectures uniquely compatible with edge-native microprocessors for field deployment.

Despite these advantages, standard deep learning models face limitations during physical deployment. Purely statistical optimization leaves them structurally blind to underlying physical laws and constraints, thereby prone to failure with noisy experimental data[26], underscoring the simulation-to-reality (sim-to-real) gap[27]. While training models exclusively on clean, synthetic data enables rapid optimization, it fundamentally biases the neural network toward idealities that do not exist in practice[18]. Conversely, training a network entirely on experimental measurements is practically unviable as real-world datasets are inherently imperfect and sparse due to the environmental constraints involved in acquiring high-precision, high-density datasets. Relying solely on these limited, sparse physical trials leads to severe underfitting, leaving the algorithm blind when confronted with changing field dynamics[26,28]. Recent efforts have demonstrated the viability of machine-learning-assisted magnetometry[26,29]. While highly effective, these approaches rely on purely statistical training, which necessitates massive, high-density datasets that can be challenging to acquire in experimental settings.

Here, we introduce a physics-guided machine learning framework optimized for ensemble NV magnetometers. By directly embedding the deterministic Zeeman splitting relationship into the training pipeline, our model enforces rapid algorithmic convergence and significantly enhances predictive accuracy. Furthermore, our hybrid approach circumvents physical data scarcity by combining scalable synthetic data generation with sparse empirical measurements, while simultaneously integrating real-world hardware non-idealities into the training pipeline.

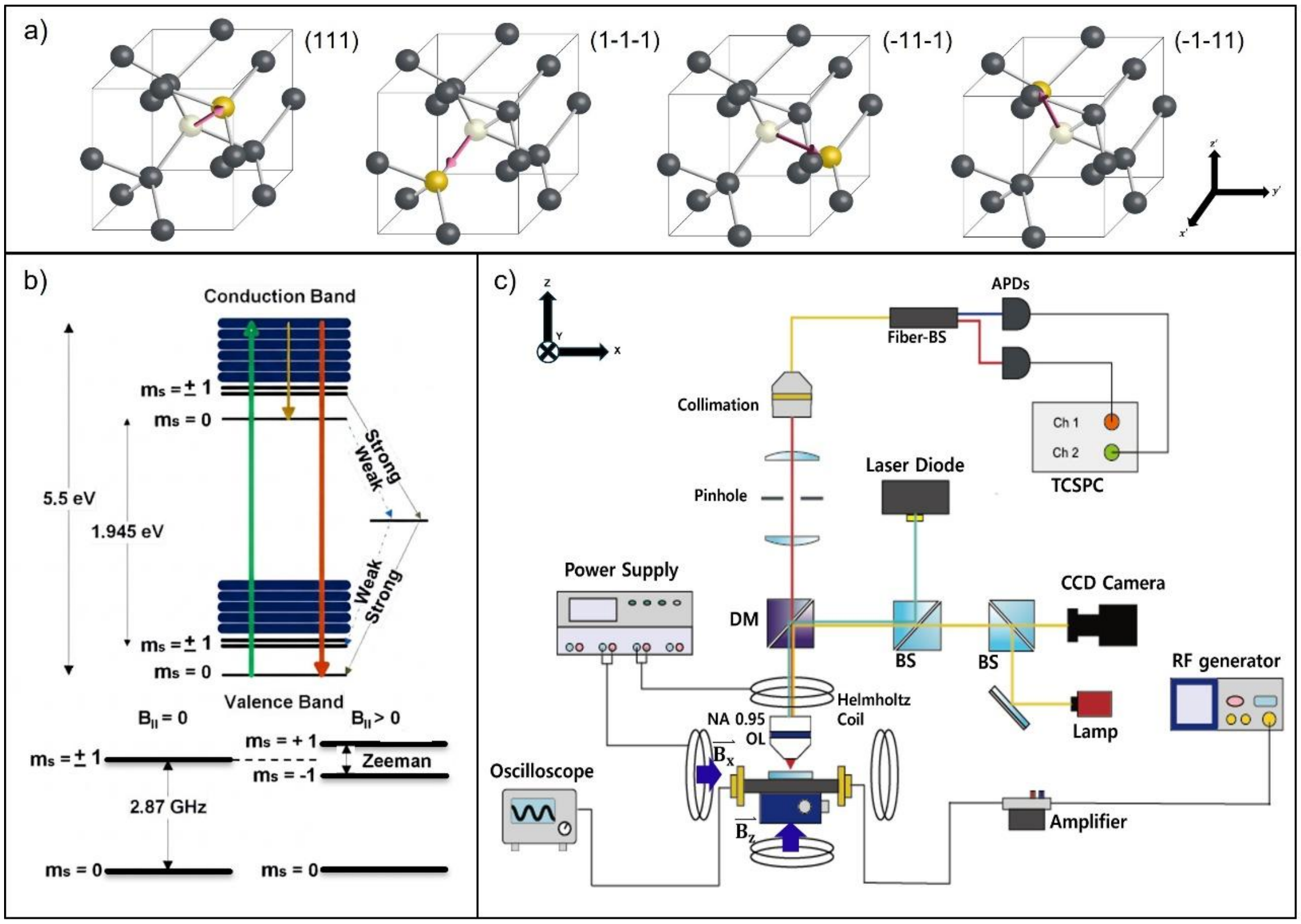

Figure 1: NV Center four-axis crystallography, energy diagram and ODMR experimental setup. (a) Atomic structure of the NV center in the diamond lattice, showing four crystallographic NV orientations. (b) Energy diagram of $NV^-$ center (c) ODMR Experimental setup.

The NV center is a point defect formed by a substitutional nitrogen atom adjacent to a lattice vacancy (Fig. 1a), whose symmetry axis aligns along one of four tetrahedrally symmetric crystallographic directions [111], [$1\bar{1}\bar{1}$], [$\bar{1}1\bar{1}$], and [$\bar{1}\bar{1}1$][30]. The negatively charged defect ($NV^{-}$) forms a spin triplet ($S$ = 1) in the ground state and it is the magnetically active state relevant for ODMR. The electron spin-spin interactions within the ground state define the spin quantization axis along the NV symmetry axis, resulting in a zero-field splitting of 2.87 GHz between the $m_s$ = 0 and $m_s$ = ±1 sublevels. When optically excited, the defect emits photoluminescence (PL) with a zero-phonon line at 637 nm. From the $m_s$= ±1 spin states, the NV center preferentially relaxes through non-radiative metastable singlet states, whereas the $m_s$= 0 state decays radiatively. This difference in relaxation pathways enables optical spin initialization into the $m_s$= 0 state and spin-state-dependent fluorescence readout, where reduced PL intensity indicates population in the $m_s$ = ±1 states. The system's ground state is governed by the Spin Hamiltonian:

$$\frac{\hat{H}}{h} = D\left(\hat{S}_z^2 - \frac{1}{3}S(S+1)\right) + E\left(\hat{S}_x^2 - \hat{S}_y^2\right) + \gamma_{NV} B \cdot \hat{S} \qquad \text{(Eq.1)}$$

To ensure dimensional consistency, the Hamiltonian is represented in frequency units $\left(\hat{H}/h\right)$, where S is the dimensionless spin-1 operator, $D$ and $E$ are the axial zero-field and transverse strain/electric field-induced splittings, and $\gamma_{NV}B$ is the Zeeman term in frequency units. Applying a microwave field drives transitions between $m_s$= 0 and $m_s$ =±1, producing a characteristic ODMR dip in PL intensity (Fig. 1b)[30].

In the presence of an external magnetic field, the $m_s$ = ±1 degeneracy is lifted via Zeeman splitting into two frequencies. In the regime where transverse-field mixing and strain-induced splitting can be neglected, the axial projection sensed by the $i^{th}$ NV family is extracted directly from the ODMR frequency separation, $\Delta\nu^{(i)} = f_+{}^{(i)} - f_-{}^{(i)}$:

$$B_{\parallel}{}^{(i)} = \frac{f_+^{(i)} - f_-^{(i)}}{2\gamma_{NV}} = \frac{\Delta\nu^{(i)}}{2\gamma_{NV}} \qquad \text{(Eq.2)}$$

where $\gamma_{NV} = g\mu_B/h$ = 28 GHz/T (2.8 MHz/G) is the electron gyromagnetic ratio. It quantifies the Zeeman coupling between the NV electron spin and the external magnetic field, defining the shift of the spin sublevels per unit field ($\Delta f = D \pm \gamma_{NV} B_{\parallel}$). Since the four NV axes are mutually non-coplanar and span three-dimensional space, simultaneous acquisition of ODMR spectra from all four families enables full reconstruction of the three-dimensional magnetic field vector $\vec{B} = (\vec{B}_x, \vec{B}_y, \vec{B}_z)$, without mechanical reorientation of the sample[31,32].

To relate this to the laboratory reference frame (x, y, z), we define local coordinate systems for the four non-coplanar NV crystallographic axes (x', y', z'), where z' represents the primary quantization axis for a given NV family. The outermost spectral resonances correspond to the specific NV family whose axis yields the largest absolute magnetic field projection ($B_{\parallel,max}$) rather than universally representing the laboratory z-axis. It should be noted that relating this multi-peak spectral span to the field projection via Equation 2 relies on the linear Zeeman approximation, assuming the applied field is strong enough to dominate local strain variations yet weak enough that transverse-field mixing remains negligible. However, as described later, our synthetic training dataset is generated using the exact diagonalization of the full ground-state spin Hamiltonian, allowing the neural network to inherently compensate for these transverse and strain non-linearities. Moreover, during synthetic data generation, the field orientation ($\theta'_B, \phi'_B$) is held constant while only the total magnitude $B_0$ is swept. This ensures the field proportionally projects across all four NV crystallographic axes simultaneously, preserving the physical symmetry of the ensemble and preventing parameter drift. Because the sample is a macroscopic ensemble, the simulation tool performs a full multi-parameter fit of the ODMR spectrum to deduce the exact polar ($\theta'_B$) and azimuthal ($\phi'_B$) angles of the magnetic field. From a tetrahedrally symmetric arrangement of the four NV orientations, their axial magnetic-field projections collectively contain information about the three-dimensional magnetic field, allowing the total magnetic-field magnitude to be directly estimated from the measured ODMR splittings. From the total magnitude ($B_0$) and these orientation angles, the full three-dimensional magnetic field components based on crystallographic axes (x', y', z') are inherently resolved: (Eq.3)

$$B_{x\prime} = B_0 \sin(\theta'_B) \cos(\phi'_B)$$
$$B_{y\prime} = B_0 \sin(\theta'_B) \sin(\phi'_B)$$
$$B_{z\prime} = B_0 \cos(\theta'_B)$$
$$B_0 = \sqrt{B_x^2 + B_y^2 + B_z^2} = \sqrt{B_{x'}^2 + B_{y\prime}^2 + B_{z\prime}^2} = \frac{\sqrt{3}}{4\gamma_{NV}} \sqrt{\Delta\nu_1^2 + \Delta\nu_2^2 + \Delta\nu_3^2 + \Delta\nu_4^2}$$

Equation 3 is valid within the linear Zeeman approximation and assumes that the resonance pairs associated with all four NV orientations can be identified. In the present study, the applied magnetic field and the diamond crystal are randomly oriented, enabling us to estimate the magnetic-field magnitude based on the ensemble ODMR spectrum. Nevertheless, with appropriate crystallographic alignment or orientation calibration and identification of the individual NV families, the approach can be extended to three-dimensional vector magnetometry[4] .

The experimental setup (Fig. 1c) is based on a confocal optical path designed for ODMR-based NV magnetometry. A 532 nm continuous-wave laser is used for excitation and optical spin initialization of the NV centers in the diamond sample. The laser beam is directed along a free-space beam path and reflected by a dichroic mirror (532 DM) toward the diamond sample. The emitted NV photoluminescence (PL) is collected in reflection geometry and coupled into fiber via a fiber collimator. The collected PL is guided to the photon counting avalanche photodiodes (APD, Excelitas SPCM-NIR), where photon arrival events are registered. Photon timing is recorded by a time-correlated single-photon counting (TCSPC) module (PicoQuant PicoHarp 300), enabling time-resolved PL measurements accessible through the acquisition software. Microwave excitation is delivered to the diamond sample via a wire antenna directly contacted to the sample surface in a pre-wired configuration. A microwave signal generated by a vector signal generator (Rohde & Schwarz SMBV 100A) is applied near 2.87 GHz to drive spin transitions between the $m_s$= 0 and $m_s$= ±1 sublevels, producing the characteristic ODMR dip in the PL intensity.

To lift the degeneracy of the four NV orientation families and resolve ensemble Zeeman split resonances, a static magnetic field is applied using a custom-designed three-axis Helmholtz coil assembly, powered by a programmable DC power supply (Rohde & Schwarz HMP4040). A controlled DC current supplied to each coil axis generates a uniform magnetic field at the sample position. The resulting ODMR spectrum, comprising two resonances per NV family, is used to extract the axial field projections and reconstruct the full three-dimensional magnetic field vector[32]. The reference magnetic field ($B_0$) was independently measured at the sample position using a Hall-effect magnetic-field probe (TD8620, 2000 G range, 0.1 G resolution). To ensure a robust dataset, the sample position and measurement location were continuously varied. While acquiring the ODMR spectra, a Helmholtz coil was used to intentionally change not only the magnitude of the B field, but also its direction. At each condition, $B_x, B_y,$ and $B_z$ were each measured 10 times, and the total field was calculated as $B_0 = \sqrt{B_x^2 + B_y^2 + B_z^2}$ in Eq.3. Because the probe was manually positioned at the sample location, this introduced unavoidable measurement-to-measurement positioning errors, which serve as the primary source of uncertainty in the reference labels. A detailed analysis of these measurement uncertainties is provided in Supplementary S3.

To evaluate the architectural optimization strategies, we first generated an idealized synthetic dataset by solving the spin Hamiltonian in Eq. 1 to extract spin resonances with fixed initial parameters such as polar and azimuthal angles, yielding clean, symmetric, multi-peak ODMR profiles without empirical hardware noise. Using this idealized dataset, we evaluated three optimization strategies implementing a shallow deep learning model utilizing K-fold cross validation (Fig. 2). The baseline methodology employs a purely statistical approach, treating the ODMR spectra as 1D vector, 2,561 pts, 1 label ($B_0$) input matrix. The architecture utilizes two hidden layers (512 and 128 units), employing Rectified Linear Unit (ReLU) activation functions and a 30% dropout rate and K-fold cross validation to mitigate overfitting. This strategy relies solely on minimizing the empirical loss between predicted and observed spectral features, without adherence to the governing physical constraints. In contrast, our physics-guided framework embeds the end-to-end frequency separation of the ODMR spectral envelope $\Delta\nu$, directly into the learning architecture. By utilizing the peak-finding methods to isolate the resonance lineshapes and fitting them to the deterministic Zeeman splitting relation, $\Delta\nu_{max} \approx 2\gamma_{NV}B_{\parallel,max}$, each spectrum is assigned a physics-guided label. This strategy successfully restricts the search space of the optimization manifold, effectively acting as a regularization[33,34]. Moreover, our ensemble neural network consists of feeding the outputs of the two models into an additional machine learning layer to yield the final predictions. In the next section, we compare the results of these optimization strategies in detail.

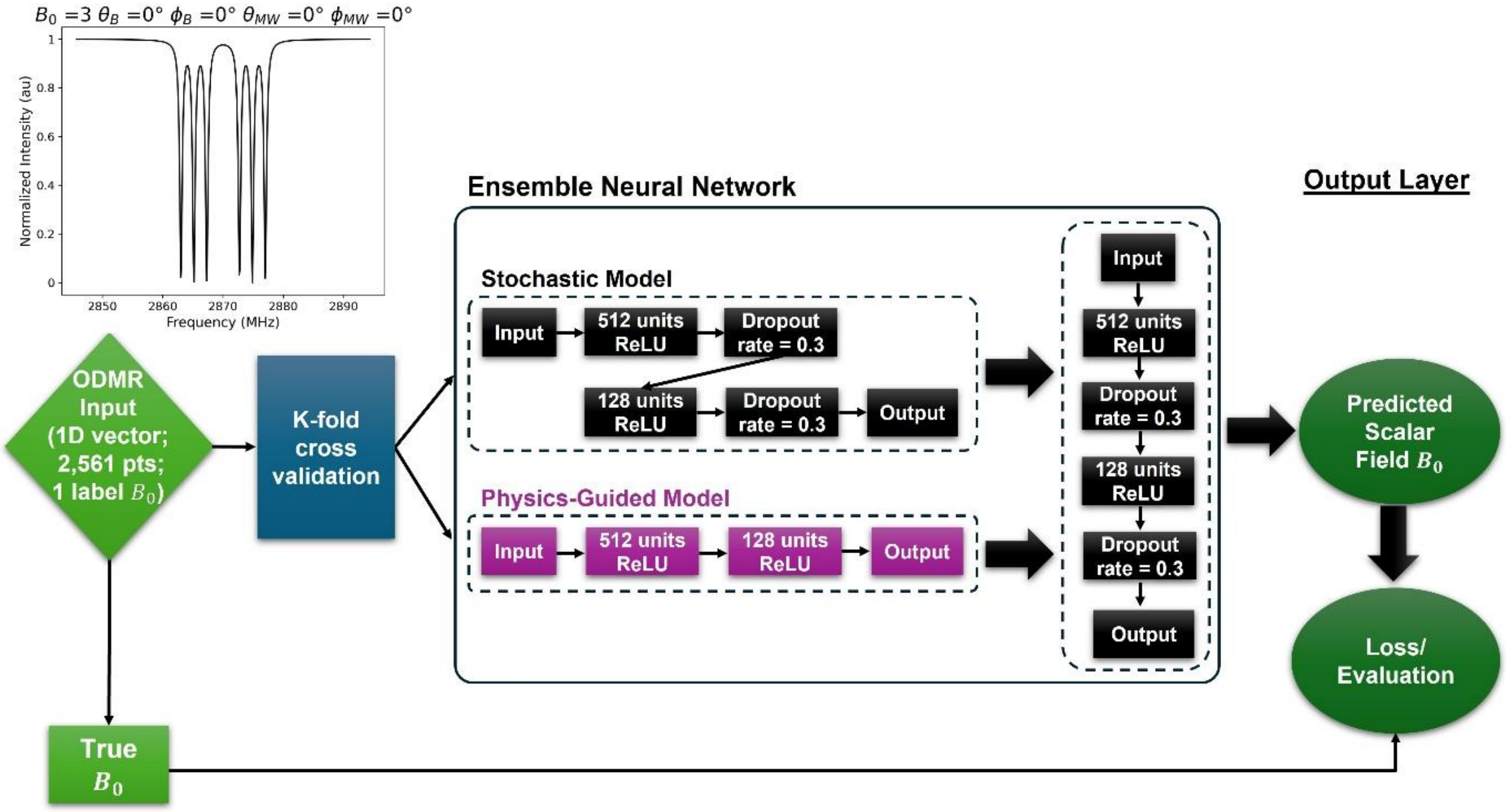

Figure 2: Structural Architecture and workflow. The pipeline takes a simulated cw ODMR spectrum (1D vector, 2,561 pts and 1 label ($B_0$)) as an input array and processes it alongside the true scalar magnetic field ($B_0$). The dataset undergoes a K-fold cross-validation branching into stochastic and physics-guided models. The outputs of the two models are fed into another layer of stochastic model forming an ensemble neural network, yielding the final outputs.

A quantitative comparison of performance metrics and error distributions among the stochastic model, the physics-guided model, and the ensemble architecture is illustrated in Figure 3. The scatter plots in the first column (Figs. 3a–c) map actual versus predicted outputs across 1,000 independent validation samples. While these initial scatter plots display the general predictive alignment for all three models, the severe limitations of purely statistical mapping are explicitly exposed in the training dynamics. Mean Squared Error (MSE) convergence profiles across a 5-fold cross-validation (Figs. 3d–f) reveal the stochastic model's severe volatility and its inability to settle into a unified global minimum. Conversely, both the physics-guided and ensemble models achieve immediate, synchronized stability within a few epochs. By directly embedding the end-to-end frequency separation of the ODMR envelope into the learning pipeline, the physics-guided architecture introduces a deterministic constraint governed by the electronic gyromagnetic ratio, $\gamma_{NV}$, suppressing spurious local minima and forcing the optimization path onto a stable, physically consistent minimum without adding backpropagation complexity.

This architectural evolution further manifests in the Mean Error ($\Delta = B_{pred.} - B_{true}$), standard deviation ($\sigma$), and localized error distribution histograms shown in the third column (Figs. 3g–i). The stochastic model (Fig. 3g) relies entirely on unconstrained Stochastic Gradient Descent (SGD) to map arbitrary numerical patterns from the raw input distribution. In the low magnetic field regime (B < 1 G), the stochastic model experiences catastrophic tracking failure, exhibiting massive, unstable residual swings. By comparison, the physics-guided framework (Fig. 3h) dramatically compresses the Mean Error down to $\Delta = 0.00142$ G. Compared to the stochastic model, this is a 99.25% reduction in the Mean Error ($\Delta$ = 0.189 G → 0.00142 G), which effectively eliminates systematic prediction bias while tightening the standard deviation to $\sigma = 0.0164$ G and smoothing out the low-field divergence.

Finally, the ensemble model (Fig. 3i) provides further refinement, compressing the Mean Error to $\Delta = 0.00051$ G and restricting the histogram spread to $\sigma = 0.0153$ G. Moving from the physics-guided baseline to this multi-channel architecture yields an additional 64.08% reduction in error, proving that it effectively screens out remaining statistical fluctuations. As a result, the ensemble model achieves a nearly 3-fold improvement in overall precision (reducing the standard deviation from 0.0454 G to 0.0153 G) and yields a 99.73% cumulative reduction in systematic bias (a 372-fold improvement in Mean Error) compared to the baseline stochastic approach.

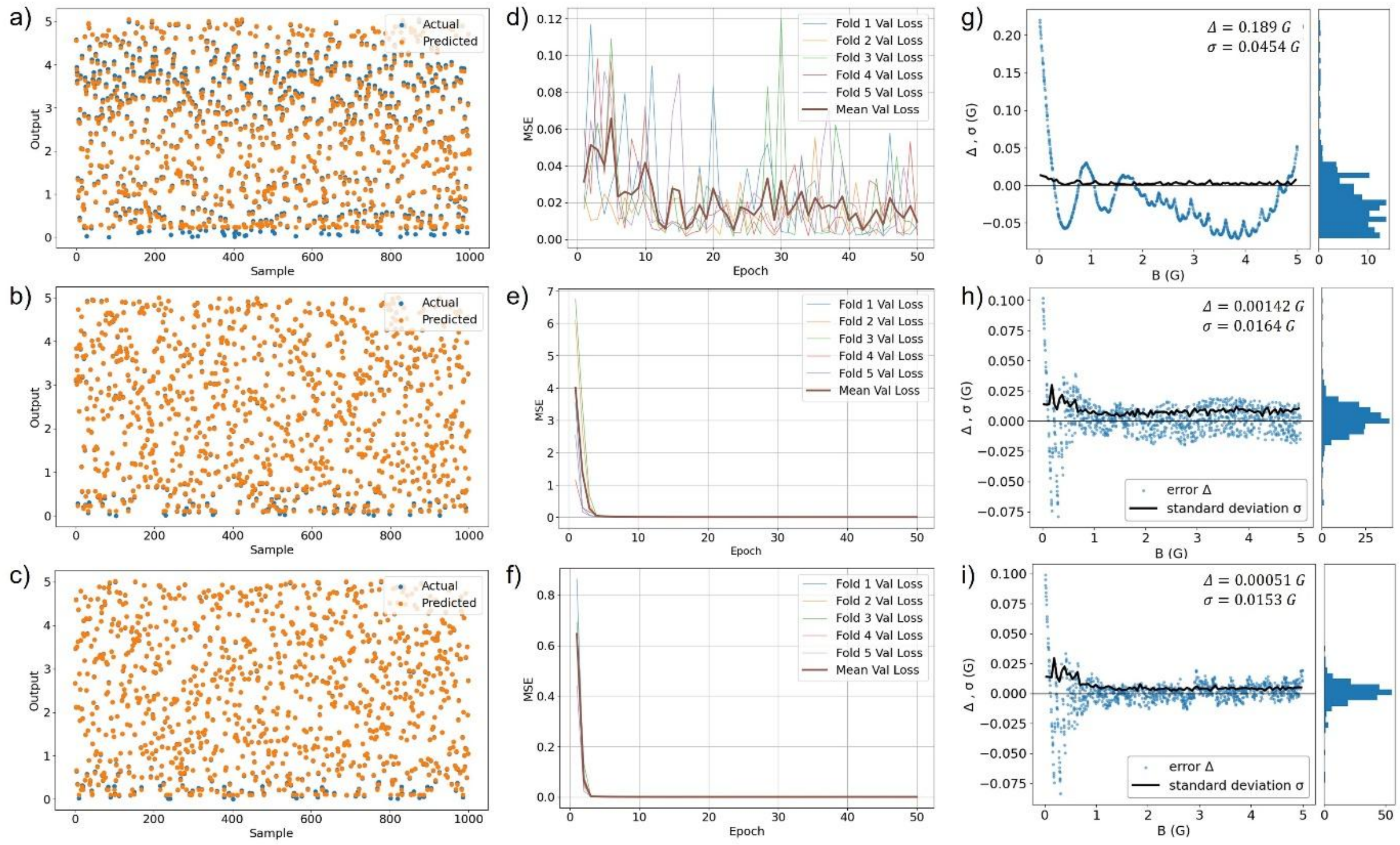

FIGURE 3: Quantitative comparison of performance across the three evaluated network architectures: stochastic (top row), physics-guided (middle row), and ensemble model (bottom row). (a) - (c) Actual-versus-predicted output scatter plots mapped as a function of sample index for 1,000 independent evaluation samples. (d) - (f) Mean squared error (MSE) optimization profiles tracking validation loss across a 5-fold cross-validation split over 50 training epochs. (g), (h), (i) Mean Error ($\Delta$) and standard deviations ($\sigma$) plotted as a function of the magnetic field, paired with localized error frequency histograms. Note: The legend in panels (h) and (i) identifying the error and standard deviation boundaries applies equally to the data plotted in panel (g).

To accurately quantify the overall prediction accuracy and account for residual cancellation, Mean Absolute Error (MAE) and Root Mean Square Error (RMSE) are reported alongside these metrics in Table 1. Furthermore, to evaluate edge-deployment viability, Table 1 details the computational execution times. These benchmarks were recorded on a standard commercial workstation (AMD Ryzen 5 5600X CPU, 32 GB RAM, NVIDIA GeForce RTX 3050 GPU), emphasizing that high-performance computing clusters are not required for model training or deployment. Even on this standard hardware, all three models exhibit sub-4 millisecond prediction times per spectrum, confirming that this shallow architecture bypasses the severe computational overhead associated with conventional iterative solvers. While increasing the input data density could theoretically improve interpolation for a purely statistical approach, it would also exacerbate training inefficiencies and increase the computational overhead. Furthermore, in data-sparse regimes, such a strategy typically fails to improve generalization, as the model remains susceptible to capturing non-physical correlations within the spectral noise floor[35].

| Model | Mean Error $\Delta$ (G) | Standard Deviation $\sigma$ (G) | MAE (G) | RMSE (G) | Precision Improvement Factor | Bias Reduction Factor | Training Time (s) | Prediction Time (ms) |
|---|---|---|---|---|---|---|---|---|
| Stochastic | 0.189 | $\pm$0.0454 | 0.028 | 0.194 | 1.0 | 1 | 35.0 | 1.03 |
| Physics-Guided | 0.00142 | $\pm$0.0164 | 0.011 | 0.0164 | 2.8 | 133 | 25.2 | 0.83 |
| Ensemble | 0.00051 | $\pm$0.0153 | 0.009 | 0.0153 | 3.0 | 372 | 22.0 | 3.22 |

Table 1: Performance summary for the three evaluated neural network architectures, including accuracy metrics and execution times.

Directly deploying our synthetically optimized neural network onto experimentally measured ODMR data initially failed to yield accurate predictions. Lacking real-world non-idealities, such as shot noise, power broadening, and optical asymmetries, the model faces a severe sim-to-real mismatch[26,29]. Training models for physical systems presents a fundamental data conundrum: robust generalization requires massive datasets, yet acquiring high-density experimental measurements is resource-prohibitive, and relying on too few data points causes severe underfitting. Our hybrid approach circumvents this bottleneck by utilizing 49 sparse experimental ODMR spectra as empirical seeds to initialize a parameterized simulation engine (Supplement S1-1 and S1-2). During this step, primary hardware non-idealities are accounted for: the microwave field vector orientation is tuned to match empirical contrast differences and peak asymmetries, and Gaussian shot noise variance is scaled to the experimental noise floor. After locking in these hardware-specific baselines, the engine scales data generation by sweeping only the scalar magnetic field $B_0$. This produces a comprehensive hybrid dataset of 19,419 spectra that explicitly embeds real-world complexities, preventing the network from developing a bias toward idealized data and forcing it to adapt directly to actual field dynamics.

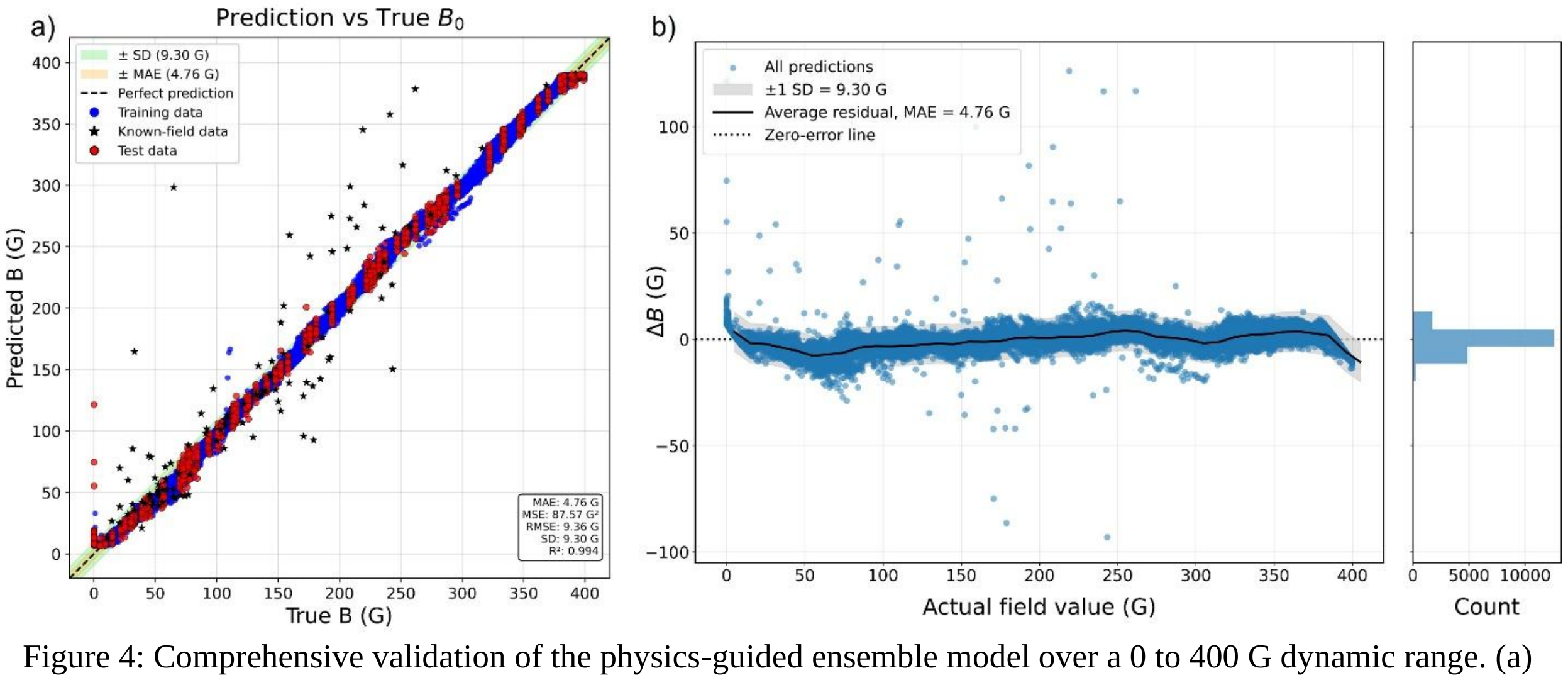

Figure 4: Comprehensive validation of the physics-guided ensemble model over a 0 to 400 G dynamic range. (a) Parity plot comparing predicted versus true scalar magnetic field ($B_0$). The plot overlays the hybrid training dataset (blue circles), the unseen 20% hybrid test set (red circles), and the raw, uncalibrated experimental hardware data

(black stars). The framework achieves an overall MAE of 4.76 G, a standard deviation of $\pm$ 9.30 G, and an $R^2$ of 0.994. (b) Residual error ($\Delta B$) plotted as a function of the actual field value, demonstrating strong low-field convergence and stable sim-to-real transfer across the full measurement spectrum.

To fully assess the comprehensive performance and sim-to-real transferability of the physics-guided ensemble model, the framework was deployed across 15,540 hybrid training spectra (blue circles), 3,879 hybrid test spectra (red circles), and 108 raw experimental test spectra (black stars). Crucially, all test sets were strictly excluded from the training phase. Figure 4 illustrates these predictions spanning a broad 0 to 400 G dynamic range. As shown in the parity plot (Fig. 4a), the hybrid training data tightly converges along the identity line, confirming the architecture successfully learned the underlying physical relationships. Furthermore, predictions on the completely unseen hybrid test set, representing a 20% holdout, demonstrate robust convergence. Moreover, our model successfully maintains predictive accuracy when deployed on 108 raw, uncalibrated experimental spectra. While prediction variance naturally broadens in the high-field regime, which is consistent with growing experimental uncertainty and signal degradation at elevated fields (Supplementary S3), the model exhibits strong convergence at lower field values, proving competitive applicability across the entire spectrum.

Despite the broadened range (400 G), the overall error metrics remain low, achieving a Mean Absolute Error (MAE) of 4.76 G and a standard deviation (SD) of $\pm$ 9.30 G. This accuracy is quantitatively supported by a coefficient of determination ($R^2$) of 0.994, indicating that statistically less than 0.6% of the variance in the evaluated data remains unexplainable. The corresponding residual analysis (Fig. 4b) further confirms this structural stability. Ultimately, the evaluation of the complete dataset shows that the physics-guided framework performs successfully on noisy, real-world hardware data using only 49 experimental seeds. This proven sim-to-real bridge provides strong motivation that further scaling the physics-guided hybrid data generation may yield even higher precision for edge-deployed sensors across all field ranges.

Beyond initial calibration, this hybrid framework is exceptionally well-suited for long-term field deployment. During normal operation, the model does not require continuous retraining. Retraining is only necessary if the sensor experiences a fundamental physical change, such as severe thermal drift or structural strain that alters the ODMR signatures beyond the hardware non-idealities already accounted for in the initial training set. In such events, an exhaustive experimental dataset is not required. Acquiring a few new sparse measurements and passing them through the simulation rapidly generates a fresh synthetic calibration set to update the network, drastically minimizing the experimental overhead required to manage drift over time.

In summary, we have demonstrated a physics-guided hybrid machine learning framework for ensemble NV diamond magnetometers. In the synthetic ablation study, embedding the Zeeman splitting relation directly into the learning architecture enabled rapid, stable convergence, yielding a nearly 3-fold improvement in precision and a 372-fold reduction in systematic bias over a purely statistical baseline. To bridge the sim-to-real gap without requiring exhaustive physical datasets, we anchored a parameterized simulation engine with just 49 sparse experimental ODMR spectra to lock in hardware non-idealities and generate over 19,000 hybrid spectra across a 0 to 400 G dynamic range. Evaluated across this comprehensive dataset alongside 108 raw, uncalibrated experimental spectra, the framework maintained robust predictive accuracy ($MAE = 4.76\,G,\ \sigma = \pm\,9.30\,G,$ and $R^2 = 0.994$), confirming the model's successful performance on noisy real-world measurements. Moreover, achieving sub-4 ms inference times per spectrum with minimal computational overhead on standard commercial hardware demonstrates strong viability for real-time edge deployment. Ultimately, this architecture extends beyond NV magnetometry, offering a scalable, generalizable paradigm for diverse physical sensing platforms where acquiring high-density experimental data remains a critical bottleneck.

## SUPPLEMENTARY MATERIAL

See supplementary material for the complete list of physical parameters and the machine learning architecture used in the hybrid data training.

## ACKNOWLEDGMENTS

This work was supported by the National Science Foundation (NSF) under the Centers of Research Excellence in Science and Technology (CREST) program (Grant No. HRD-1914777) and the IITP (Institute of Information & Communications Technology Planning & Evaluation) grant (RS-2025-25464832).

## CONFLICT OF INTEREST

The authors have no conflicts to disclose.

## AUTHOR CONTRIBUTIONS

Jonathan Daniel and Sangwoo Lee contributed equally to this work.

## DATA AVAILABILITY STATEMENT

The code, data, trained models, and documentation supporting the findings of this study will be publicly available upon publication at: https://github.com/j-daniel-csusb/nv-odmr-simulation.
Materials are available to reviewers during peer review as Supplementary Material for Review Only.

[1] S. Hong, M.S. Grinolds, L.M. Pham, D. Le Sage, L. Luan, R.L. Walsworth, and A. Yacoby, "Nanoscale magnetometry with NV centers in diamond," MRS Bull. **38**(2), 155–161 (2013).

[2] L. Rondin, J.-P. Tetienne, P. Spinicelli, C. Dal Savio, K. Karrai, G. Dantelle, A. Thiaville, S. Rohart, J.-F. Roch, and V. Jacques, "Nanoscale magnetic field mapping with a single spin scanning probe magnetometer," Appl. Phys. Lett. **100**(15), 153118 (2012).

[3] T.X. Zhou, R.J. Stöhr, and A. Yacoby, "Scanning diamond NV center probes compatible with conventional AFM technology," Appl. Phys. Lett. **111**(16), 163106 (2017).

[4] H. Clevenson, L.M. Pham, C. Teale, K. Johnson, D. Englund, and D. Braje, "Robust high-dynamic-range vector magnetometry with nitrogen-vacancy centers in diamond," Appl. Phys. Lett. **112**(25), 252406 (2018).

[5] F.M. Stürner, A. Brenneis, T. Buck, J. Kassel, R. Rölver, T. Fuchs, A. Savitsky, D. Suter, J. Grimmel, S. Hengesbach, M. Förtsch, K. Nakamura, H. Sumiya, S. Onoda, J. Isoya, and F. Jelezko, "Integrated and Portable Magnetometer Based on Nitrogen-Vacancy Ensembles in Diamond," Adv. Quantum Technol. **4**(4), 2000111 (2021).

[6] Y. Xie, H. Yu, Y. Zhu, X. Qin, X. Rong, C.-K. Duan, and J. Du, "A hybrid magnetometer towards femtotesla sensitivity under ambient conditions," Sci. Bull. (Beijing). **66**(2), 127–132 (2021).

[7] G. Balasubramanian, I.Y. Chan, R. Kolesov, M. Al-Hmoud, J. Tisler, C. Shin, C. Kim, A. Wojcik, P.R. Hemmer, A. Krueger, T. Hanke, A. Leitenstorfer, R. Bratschitsch, F. Jelezko, and J. Wrachtrup, "Nanoscale imaging magnetometry with diamond spins under ambient conditions," Nature **455**(7213), 648–651 (2008).

[8] J.R. Maze, P.L. Stanwix, J.S. Hodges, S. Hong, J.M. Taylor, P. Cappellaro, L. Jiang, M.V.G. Dutt, E. Togan, A.S. Zibrov, A. Yacoby, R.L. Walsworth, and M.D. Lukin, "Nanoscale magnetic sensing with an individual electronic spin in diamond," Nature **455**(7213), 644–647 (2008).

[9] J.F. Barry, J.M. Schloss, E. Bauch, M.J. Turner, C.A. Hart, L.M. Pham, and R.L. Walsworth, "Sensitivity optimization for NV-diamond magnetometry," Rev. Mod. Phys. **92**(1), 15004 (2020).

[10] B. Zhao, H. Guo, R. Zhao, F. Du, Z. Li, L. Wang, D. Wu, Y. Chen, J. Tang, and J. Liu, "High-Sensitivity Three-Axis Vector Magnetometry Using Electron Spin Ensembles in Single-Crystal Diamond," IEEE Magn. Lett. **10**, 1–4 (2019).

[11] J.M. Schloss, J.F. Barry, M.J. Turner, and R.L. Walsworth, "Simultaneous Broadband Vector Magnetometry Using Solid-State Spins," Phys. Rev. Appl. **10**(3), 34044 (2018).

[12] B. Chen, X. Hou, F. Ge, X. Zhang, Y. Ji, H. Li, P. Qian, Y. Wang, N. Xu, and J. Du, "Calibration-Free Vector Magnetometry Using Nitrogen-Vacancy Center in Diamond Integrated with Optical Vortex Beam," Nano Lett. **20**(11), 8267–8272 (2020).

[13] M. Tsukamoto, K. Ogawa, H. Ozawa, T. Iwasaki, M. Hatano, K. Sasaki, and K. Kobayashi, "Vector magnetometry using perfectly aligned nitrogen-vacancy center ensemble in diamond," Appl. Phys. Lett. **118**(26), 264002 (2021).

[14] R.S. Cassel, W.G. Tobias, and B.L. Marlow, "PR-23-0577-Quantum vs. Classical Complementary PNT," MITRE, (2023).

[15] M. Simanovskaia, K. Jensen, A. Jarmola, K. Aulenbacher, N. Manson, and D. Budker, "Sidebands in optically detected magnetic resonance signals of nitrogen vacancy centers in diamond," Phys. Rev. B **87**(22), 224106 (2013).

[16] A. Jarmola, Z. Bodrog, P. Kehayias, M. Markham, J. Hall, D.J. Twitchen, V.M. Acosta, A. Gali, and D. Budker, "Optically detected magnetic resonances of nitrogen-vacancy ensembles in 13C-enriched diamond," Phys. Rev. B **94**(9), 94108 (2016).

[17] M. Krenn, J. Landgraf, T. Foesel and F. Marquardt, "Artificial intelligence and machine learning for quantum technologies," Phys. Rev. A **107**(1), 10101 (2023).

[18] V. Cimini, I. Gianani, N. Spagnolo, F. Leccese, F. Sciarrino, and M. Barbieri, "Calibration of Quantum Sensors by Neural Networks," Phys. Rev. Lett. **123**(23), 230502 (2019).

[19] V. Dunjko and H.J. Briegel, "Machine learning & artificial intelligence in the quantum domain: a review of recent progress," Rep. Prog. Phys. **81**(7), 074001 (2018).

[20] K. Hayashi, Y. Matsuzaki, T. Taniguchi, T. Shimo-Oka, I. Nakamura, S. Onoda, T. Ohshima, H. Morishita, M. Fujiwara, S. Saito, and N. Mizuochi, "Optimization of Temperature Sensitivity Using the Optically Detected Magnetic-Resonance Spectrum of a Nitrogen-Vacancy Center Ensemble," Phys. Rev. Appl. **10**(3), 34009 (2018).

[21] M. Fujiwara, A. Dohms, K. Suto, Y. Nishimura, K. Oshimi, Y. Teki, K. Cai, O. Benson, and Y. Shikano, "Real-time estimation of the optically detected magnetic resonance shift in diamond quantum thermometry toward biological applications," Phys. Rev. Res. **2**(4), 43415 (2020).

[22] P. Welter, B.A. Jósteinsson, S. Josephy, A. Wittmann, A. Morales, G. Puebla-Hellmann, and C.L. Degen, "Fast Scanning Nitrogen-Vacancy Magnetometry by Spectrum Demodulation," Phys. Rev. Appl. **19**(3), 34003 (2023).

[23] S. Dushenko, K. Ambal, and R.D. McMichael, "Sequential Bayesian Experiment Design for Optically Detected Magnetic Resonance of Nitrogen-Vacancy Centers," Phys. Rev. Appl. **14**(5), 54036 (2020).

[24] A. Youssry, S. Todd, P. Murton, M.J. Arshad, N. Werren, A. Peruzzo, and C. Bonato, "Bayesian quantum sensing using graybox machine learning," PRX Intelligence, (2026) DOI: https://doi.org/10.1103/1fbh-z7xf.

[25] G.C. Marinó, A. Petrini, D. Malchiodi, and M. Frasca, "Deep neural networks compression: A comparative survey and choice recommendations," Neurocomputing **520**, 152–170 (2023).

[26] J. Homrighausen, L. Horsthemke, J. Pogorzelski, S. Trinschek, P. Glösekötter, and M. Gregor, "Edge-Machine-Learning-Assisted Robust Magnetometer Based on Randomly Oriented NV-Ensembles in Diamond," Sensors **23**(3), 1119 (2023).

[27] D.L. Craig, H. Moon, F. Fedele, D.T. Lennon, B. van Straaten, F. Vigneau, L.C. Camenzind, D.M. Zumbühl, G.A.D. Briggs, M.A. Osborne, D. Sejdinovic, and N. Ares, "Bridging the Reality Gap in Quantum Devices with Physics-Aware Machine Learning," Phys. Rev. X **14**(1), 11001 (2024).

[28] M.C. Caro, H.-Y. Huang, N. Ezzell, J. Gibbs, A.T. Sornborger, L. Cincio, P.J. Coles, and Z. Holmes, "Out-of-distribution generalization for learning quantum dynamics," Nat. Commun. **14**(1), 3751 (2023).

[29] M. Tsukamoto, S. Ito, K. Ogawa, Y. Ashida, K. Sasaki, and K. Kobayashi, "Accurate magnetic field imaging using nanodiamond quantum sensors enhanced by machine learning," Sci. Rep. **12**(1), 13942 (2022).

[30] L. Rondin, J.-P. Tetienne, T. Hingant, J.-F. Roch, P. Maletinsky, and V. Jacques, "Magnetometry with nitrogen-vacancy defects in diamond," Rep. Prog. Phys. **77**(5), 056503 (2014).

[31] L.M. Pham, D. Le Sage, P.L. Stanwix, T.K. Yeung, D. Glenn, A. Trifonov, P. Cappellaro, P.R. Hemmer, M.D. Lukin, H. Park, A. Yacoby, and R.L. Walsworth, "Magnetic field imaging with nitrogen-vacancy ensembles," New J. Phys. **13**, 045021 (2011).

[32] S. Steinert, F. Dolde, P. Neumann, A. Aird, B. Naydenov, G. Balasubramanian, F. Jelezko, and J. Wrachtrup, "High sensitivity magnetic imaging using an array of spins in diamond," Rev. Sci. Instrum. **81**(4), 043705 (2010).

[33] J.J. Pannell, S.E. Rigby, and G. Panoutsos, "Physics-informed regularisation procedure in neural networks: An application in blast protection engineering," Int. J. Prot. Struct. **13**(3), 555–578 (2022).

[34] M.A. Nabian and H. Meidani, "Physics-Driven Regularization of Deep Neural Networks for Enhanced Engineering Design and Analysis," J. Comput. Inf. Sci. Eng. **20**(1), 011006 (2019).

[35] K. Donhauser, A. Ţifrea, M. Aerni, R. Heckel, and F. Yang, "*Interpolation Can Hurt Robust Generalization Even When There Is No Noise*" (Adv. Neural Inf. Process. Syst. 34, 2021).

| Parameter | Description | Units |
|---|---|---|
| $\theta_{MW}, \phi_{MW}$ | Polar & azimuthal angles of microwave field | deg |
| $\theta_B, \phi_B$ | Polar & azimuthal angles of external field B | deg |
| $\theta_E, \phi_E$ | Polar & azimuthal angles of strain/electric field | deg |
| $B_0$ | Magnetic field magnitude sweep | Gauss |
| $E_0$ | Magnitude of scalar E | MHz |
| Γ (Linewidth) | Full width at half maximum (Gaussian/Lorentzian) | MHz |
| Noise Mean | Gaussian shot-noise floor level | a.u. |
| Scalar | Frequency range/spectral span | MHz |
| Noise STD | Standard deviation of noise | a.u. |
| Contrast amplitude | Optical depth of spin resonance | a.u. |

**S1-1: Comprehensive summary of fitted parameters.**

A parameterized simulation engine based on the ground-state spin Hamiltonian was initialized using baseline features extracted from 36 experimental ODMR calibration spectra. The parameterized model incorporates both intrinsic quantum state interactions and realistic hardware non-idealities. Table S1 summarizes the physical, geometric, and experimental parameters fitted by the simulation engine. The angular orientations $\theta_{MW}, \phi_{MW}$, $\theta_B, \phi_B$ and $\theta_E, \phi_E$ define the projection vectors of the applied static magnetic field, microwave drive field, and local strain/electric field relative to the four NV crystallographic axes. Linewidth broadening, contrast amplitude, and shot-noise were empirically constrained to replicate the noise floor, power broadening, and resonance asymmetries of the physical testbed. Once locked, these baseline parameters served as anchors to sweep the scalar magnetic field across a dynamic range of 0 to 400 G, yielding 19,527 densely sampled synthetic spectra for neural network training and hyperparameter optimization.

| **Sample Index** | $\boldsymbol{B_0}$ **(*G*)** | $\boldsymbol{\theta_B}$ **(deg)** | $\boldsymbol{\phi_B}$ **(deg)** | Note (B-field source) |
|---|---|---|---|---|
| **1** | 0 | 0 | 0 | Zero B-field |
| **2** | 2.25 | 0 | 0 | Helmholtz only |
| **3** | 2.5 | 0 | 0 | Helmholtz only |
| **4** | 2.5 | 0 | 0 | Helmholtz only |
| **5** | 3.75 | 0 | 0 | Helmholtz only |
| **6** | 3.75 | 0 | 0 | Helmholtz only |
| **7** | 4.25 | 0 | 0 | Helmholtz only |
| **8** | 6 | 0 | 0 | Helmholtz only |
| **9** | 6 | 0 | 0 | Helmholtz only |
| **10** | 6.25 | 0 | 0 | Helmholtz only |
| **11** | 8 | 0 | 0 | Helmholtz only |
| **12** | 8 | 0 | 0 | Helmholtz only |
| **13** | 8.25 | 0 | 0 | Helmholtz only |
| **14** | 9.5 | 0 | 0 | Helmholtz only |
| **15** | 9.5 | 0 | 0 | Helmholtz only |
| **16** | 9.5 | 0 | 0 | Helmholtz only |
| **17** | 16.5 | 16 | 22 | Neodymium only |
| **18** | 18.8 | 18 | 19 | Neodymium only |
| **19** | 18.8 | 18 | 19 | Neodymium only |
| **20** | 23 | 112 | 33 | Helmholtz only |
| **21** | 29.5 | 74 | 13 | Helmholtz & Neodymium |
| **22** | 30.5 | 70 | 196 | Helmholtz & Neodymium |
| **23** | 31 | 67 | 96 | Helmholtz & Neodymium |
| **24** | 31.8 | 74 | 206 | Helmholtz & Neodymium |
| **25** | 32 | 11 | 122 | Helmholtz & Neodymium |
| **26** | 34 | 32 | 75 | Helmholtz & Neodymium |
| **27** | 36 | 20 | 107 | Helmholtz & Neodymium |
| **28** | 37 | 20 | 104 | Helmholtz & Neodymium |
| **29** | 37 | 76 | 11 | Helmholtz & Neodymium |
| **30** | 37.3 | 22 | 79 | Helmholtz & Neodymium |
| **31** | 38 | 77 | 20 | Helmholtz & Neodymium |
| **32** | 38 | 77 | 201 | Helmholtz & Neodymium |
| **33** | 38.5 | 25 | 80 | Helmholtz & Neodymium |
| **34** | 38.5 | 68 | 10 | Helmholtz & Neodymium |
| **35** | 38.8 | 77 | 203 | Helmholtz & Neodymium |
| **36** | 64.43 | 138 | 136 | |
| **37** | 69 | 236 | 45 | |
| **38** | 107 | 236 | 45 | |
| **39** | 121.01 | 37 | 1 | |
| **40** | 162.1 | 48 | 73 | |
| **41** | 162.2 | 54 | 66 | |
| **42** | 166 | 56 | 69 | |
| **43** | 171.1 | 124 | 278 | |
| **44** | 187.1 | 50 | 0 | |
| **45** | 188 | 13 | 16 | |
| **46** | 195 | 214 | 121 | |
| **47** | 205 | 215 | 119 | |
| **48** | 225.6 | 222 | 93 | |
| **49** | 295 | 236 | 45 | |

**S1-2: Summary of the raw experimental ODMR calibration spectra used to seed the hybrid dataset generation.** Each entry details the measured total magnetic field magnitude ($B_0$) its associated measurement uncertainty, and the two fitted orientation angles ($\theta_B$) and ($\phi_B$).

S1-2 presents the experimental measurements utilized to seed our hybrid data generation pipeline. For each calibration spectrum, the reference magnetic field magnitude ($B_0$) and its corresponding measurement uncertainty were independently recorded using a Hall-effect probe at the sample location. The raw ODMR spectra were subsequently parameterized using the spin Hamiltonian to extract the exact polar ($\theta_B$) and azimuthal ($\phi_B$) orientation angles of the applied external field. By locking these extracted angles alongside the empirical hardware non-idealities, we establish a fixed, realistic physical trajectory for the simulation engine. The neural network's training dataset is then massively expanded by sweeping the scalar magnitude $B_0$ along these anchored orientations, successfully bridging the gap between sparse physical measurements and the large-scale data requirements of the machine learning model.

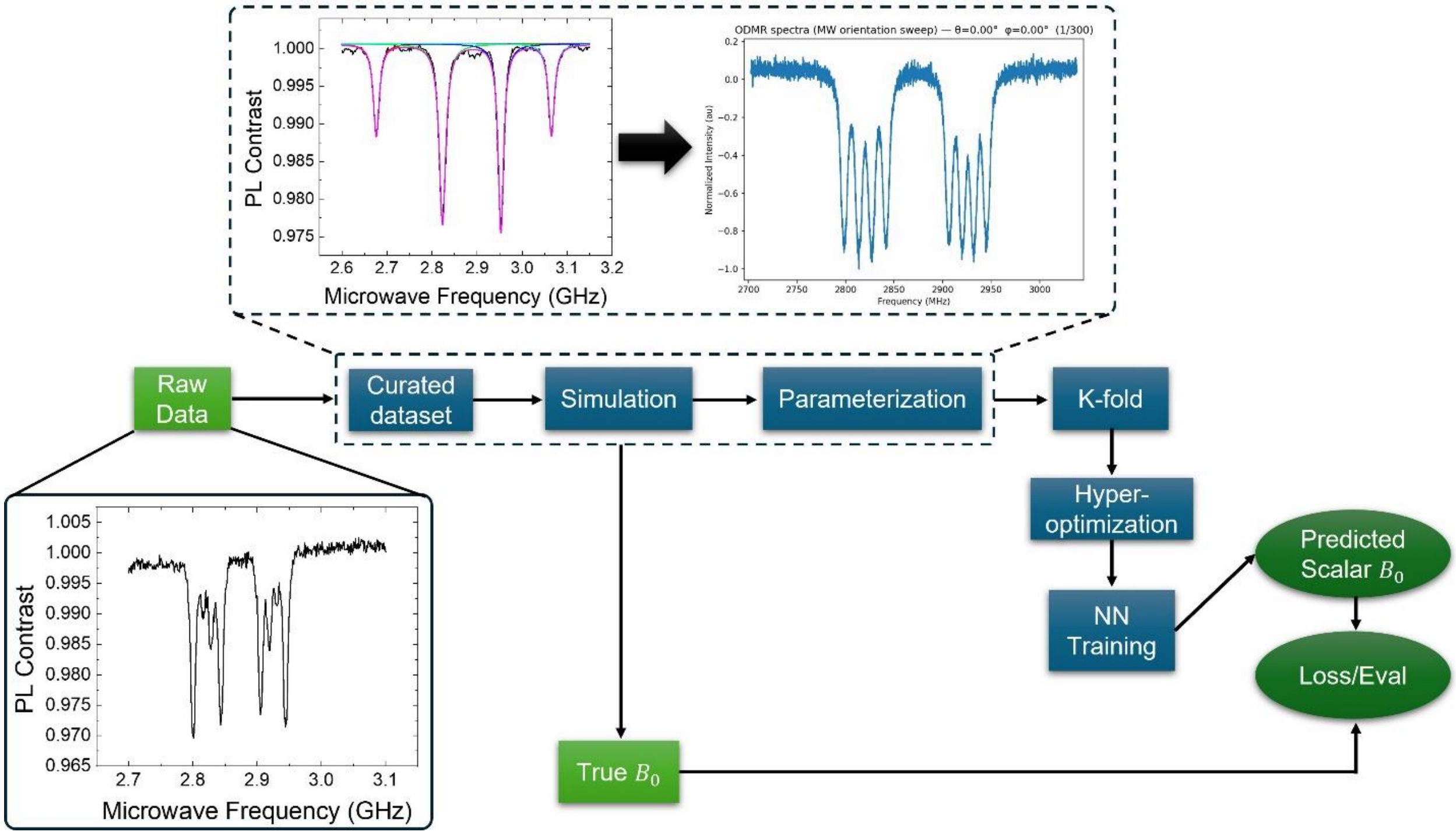

**S2: Schematic workflow and architectural pipeline of the physics-guided hybrid training framework.** Raw experimental ODMR spectra serve as empirical seeds for the simulation engine. The pipeline fits and locks baseline physical non-idealities as shown in S1 to generate a dense, parameterized dataset of 19,527 hybrid spectra. The expanded dataset undergoes K-fold cross-validation and hyperparameter optimization before final neural network training, enabling robust prediction of uncalibrated experimental $B_0$ values.

The hybrid training architecture overcomes data scarcity and experimental hardware drift by coupling empirical calibration measurements with a deterministic simulation workflow. The process initiates with raw data ingestion, where 49 unconditioned experimental ODMR spectra acquired under laboratory conditions are curated and routed to a multi-parameter fitting module. Using these locked baseline parameters, the parameterized solver sweeps the longitudinal magnetic field $B_0$, generating a dense training matrix of 19,527 synthetic spectra paired with true ground-truth labels. This populated dataset is routed through a K-fold cross-validation scheme and hyperparameter optimization pipeline to mitigate overfitting and determine optimal neural weights. During inference on real experimental results, the full ensemble network architecture is utilized. Because the neural network absorbs the physical complexities during the training phase, the raw 1D ODMR spectrum is fed directly into the model to predict the scalar field. Absolutely no peak finding, feature extraction, or spectral fitting remains necessary during prediction.

| Reference field (G) | Gaussmeter SD (G) | Relative Gaussmeter SD | Total 1mm offset | Total 2mm offset |
|---|---|---|---|---|
| 21-50 | 0.8 (0.4-1.1) | 2.2% (1.5-2.9) | ±3.9% (1.5 G) | ±7.0% (2.7 G) |
| 50-100 | 2.3 (0.8-7.0) | 3.0% (1.5-8.0) | ±4.9% (3.5 G) | ±8.5% (6.0 G) |
| 100-150 | 3.4 (1.4-6.5) | 2.6% (1.1-4.8) | ±5.4% (7.1 G) | ±10.1% (13.2 G) |
| 150-200 | 5.6 (3.1-13.7) | 3.2% (2.0-7.1) | ±6.1% (10.6 G) | ±11.2% (19.5 G) |
| 200-250 | 6.2 (1.8-15.7) | 2.7% (0.7-7.1) | ±6.3% (14.7 G) | ±12.1% (28.3 G) |
| 250-370 | 8.2 (2.7-17.3) | 2.8% (1.0-5.4) | ±6.6% (17.9 G) | ±12.7% (34.5 G) |
| All | 4.5 (0.4-17.3) | 2.8% (0.7-8.0) | | |

**S3-1: Reference uncertainty by field strength**

Tables S3-1 and S3-2 detail the estimated reference measurement uncertainties and spatial field gradients across the experimental testbed. Table S3-1 breaks down the expected reference uncertainty into two primary components: the statistical variance of the independent gaussmeter readings (based on ten repeated measurements per field condition) and the systematic error introduced by spatial displacement. While the absolute gaussmeter standard deviation (SD) scales with field magnitude, the relative SD remains largely independent of field strength, averaging 2.8% overall.

Figure S3-2 illustrates the steep spatial magnetic field gradient relative to the distance from the magnet. Because the gaussmeter sensing element is physically offset from the microscopic NV sample volume by an estimated 1 to 2 mm, this local gradient introduces significant measurement discrepancies. As shown in Table S3-1, at the highest evaluated field ranges, the combined expected reference uncertainty reaches up to ±6.6% for a 1 mm displacement and ±12.7% for a 2 mm displacement. Consequently, prediction residuals between the neural network output and the reference measurements that fall within these bounds cannot be strictly attributed to model inaccuracy, as they are fully enveloped by the physical measurement limits of the uncalibrated testbed.

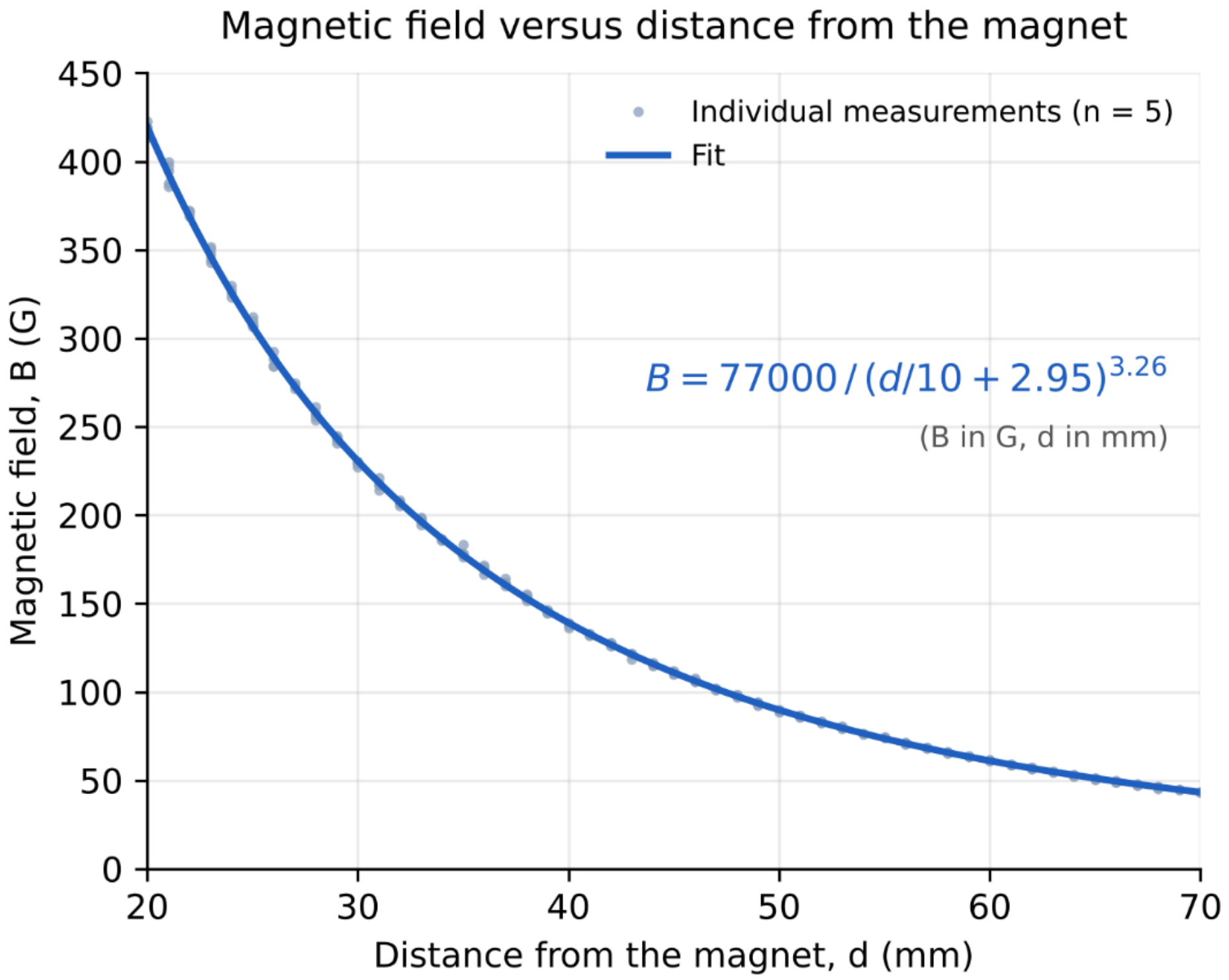

**S3-2: Magnetic field versus distance from the magnet**